\documentstyle[epsf]{mn}
\newcommand{\cld}{{\sc cloudy}}
\newcommand{\amoeba}{{\sc amoeba}}
\newcommand{\str}{Str\"omgren}
\newcommand{\iras}{{\it IRAS}}
\newcommand{\iue}{{\it IUE}}
\newcommand{\md}{\mbox{\rm d}}
\newcommand{\al}[2]{#1\,{\sc #2}}
\newcommand{\fb}[2]{[#1\,{\sc #2}]}
\newcommand{\w}{$\lambda$}
\newcommand{\jb}{{\rm\sl B}}
\newcommand{\jv}{{\rm\sl V}}
\newcommand{\ha}{\relax\ifmmode{\rm H\alpha}\else{\rm H$\alpha$}\fi}
\newcommand{\hb}{\relax\ifmmode{\rm H\beta}\else{\rm H$\beta$}\fi}
\newcommand{\hg}{\relax\ifmmode{\rm H\gamma}\else{\rm H$\gamma$}\fi}
\newcommand{\hd}{\relax\ifmmode{\rm H\delta}\else{\rm H$\delta$}\fi}
\newcommand{\he}{\relax\ifmmode{\rm H\epsilon}\else{\rm H$\epsilon$}\fi}
\newcommand{\mic}{\relax\ifmmode{\mu{\rm m}}\else{$\mu$m}\fi}
\newcommand{\zm}{\relax\ifmmode{\rm M_\odot}\else{M$_\odot$}\fi}
\newcommand{\zl}{\relax\ifmmode{\rm L_\odot}\else{L$_\odot$}\fi}
\newcommand{\kms}{\relax\ifmmode{\rm km\,s^{-1}}\else{km\,s$^{-1}$}\fi}
\newcommand{\mgn}{{\rm mag}}
\def\lg{{\rm log}}
\newcommand{\sct}{Section}

\newcommand{\tbl}{Table}
\newcommand{\sctr}[1]{\multicolumn{1}{c}{#1}}
\newcommand{\ctr}[3]{\multicolumn{#1}{c}{#2\hspace*{#3mm}}}
\newcommand{\bl}{\multicolumn{2}{c}{}}
\newcommand{\m}{\rlap{$^\dagger$}}
\newcommand{\mm}{\rlap{$^\ddagger$}}
\newcommand{\mmm}{\rlap{$^\S$}}
\newcommand{\U}{\rlap{:}}
\newcommand{\f}{\phantom{0}}
\newcommand{\x}[1]{\hspace*{#1mm}}
\def\nt#1{\vtop{\footnotesize\hsize=\columnwidth\leavevmode#1\hspace*{\fill}}}
\def\ntd#1{\vtop{\footnotesize\hsize=\textwidth\leavevmode#1\hspace*{\fill}}}
\newdimen\mbtindent
\begin{document}

\title[Photo-ionization modelling of PN -- Galactic bulge nebulae]
{Photo-ionization modelling of planetary nebulae\\
II. Galactic bulge nebulae, a comparison with literature results}
\author[P.A.M. van Hoof and G.C. Van de Steene]
{P.A.M. van Hoof$^{1,2}$ and
G.C. Van de Steene$^{3,4}$\\
$^1$Kapteyn Astronomical Institute, P.O. Box 800, 9700 AV Groningen,
The Netherlands\\
$^2$University of Kentucky, Dept.\ of Physics and Astronomy, 177 CP Building, Lexington, KY 40506--0055, USA\\
$^3$European Southern Observatory, Casilla 19001, Santiago 19, Chile\\
$^4$Research School for Astronomy and Astrophysics,
Mount Stromlo Observatory,
Private Bag, Weston Creek P.O., ACT 2611, Australia}
\date{received, accepted}
\maketitle

\begin{abstract}
We have constructed photo-ionization models of five galactic bulge planetary
nebulae using our automatic method which enables a fully self-consistent
determination of the physical parameters of a planetary nebula. The models are
constrained using the spectrum, the \iras\ and radio fluxes and the angular
diameter of the nebula. We also conducted a literature search for physical
parameters determined with classical methods for these nebulae. Comparison of
the distance independent physical parameters with published data shows that
the stellar temperatures generally are in good agreement and can be considered
reliable. The literature data for the electron temperature, electron density
and also for the abundances show a large spread, indicating that the use of
line diagnostics is not reliable and that the accuracy of these methods needs
to be improved. Comparison of the various abundance determinations indicates
that the uncertainty in the electron temperature is the main source of
uncertainty in the abundance determination. The stellar magnitudes predicted
by the photo-ionization models are in good agreement with observed values.
\end{abstract}

\begin{keywords}
methods: data analysis --
planetary nebulae: general --
planetary nebulae: abundances --
planetary nebulae: individual: H~1$-$40; M~1$-$20; M~2$-$4; M~2$-$23; M~3$-$15
\end{keywords}

\section{Introduction}
\label{intro:ch3}

In van Hoof \& Van de Steene (1999, Paper~I) we presented and tested a new
method to derive simultaneously and self-consistently all physical parameters
of a planetary nebula from a set of observed quantities. A modified version of
the photo-ionization code \cld\ (Ferland 1993) is used to calculate various
models, searching for a best fit of the predictions to the observables in an
automated way. This method uses emission line ratios, the angular diameter,
the radio and the infrared flux to constrain the model. It also takes dust
into account in the radiative transport. With this method we are able to
determine the stellar temperature and luminosity, the inner, \str\ and outer
radius of the nebula, the density, the dust-to-gas mass ratio and the
abundances. We investigated the accuracy of the determination of the physical
parameters by applying this method to an artificial set of observables. First
we proved that this method can pass a formal convergence test. Subsequently we
introduced either measurement errors in the observables or changed the model
assumptions, and investigated how this affects the best-fit model. In this way
we gained an understanding of the robustness of our method and hence of the
reliability of the physical parameters. Our method was also compared with
classical methods to determine the electron temperature and density and
nebular abundances. It was shown that our method suffers less from noise in
the spectrum than the classical line diagnostics. However, this advantage may
be lost if the model assumptions are not appropriate for the nebula being
studied. The weakest points are currently the use of a blackbody
approximation, the assumption that the inner dust radius coincides with the
inner gas radius and the assumption of spherical symmetry.

Distance determinations of Planetary Nebulae (PNe) are still very problematic.
Various methods are in use, but the range in distances obtained is often very
large and no method has found general acceptance. Reviews of the current
status can be found in Pottasch \cite{c2:rev}, Terzian \cite{c2:terzian} and
Pottasch \cite{c2:pot:dens}. The lack of a reliable, generally applicable
method to determine distances to PNe poses a problem when using
photo-ionization models. To circumvent this problem we applied the method to a
small sample of galactic bulge nebulae, which can be assumed to be all at the
same well-known distance. Our aim is to study the accuracy of the
determination of physical parameters by comparing our results with other
literature values.

A summary of the method and model assumptions is given in \sct~\ref{method}.
The sample selection is presented in \sct~\ref{sample} and the modelling
results in \sct~\ref{results}. Each PN in the sample is discussed
individually, with special emphasis on the problems encountered during the
modelling in \sct~\ref{remarks}. The resulting physical parameters are
discussed by comparing them with results from other studies in the literature
in \sct~\ref{discussion}. Our conclusions are given in \sct~\ref{conclusions}.

\section{Summary of the model assumptions and the method}
\label{method}

The model assumptions and the method were extensively described and discussed
in Paper~I. This section presents only a brief summary.

To model the planetary nebula, we use a modified version of the
photo-ionization code \cld\ 84.06 (Ferland 1993).

The model for the PN is quite simple, and comprises the following assumptions:
\begin{enumerate}
\renewcommand{\theenumi}{(\arabic{enumi})}
\item{The central star has a blackbody spectrum.}
\item{The nebula is spherically symmetric.}
\item{The density is constant inside the \str\ radius of the nebula,
and varies as $1/r^2$ outside.}
\item{Dust grains are intermixed with the gas at a constant dust-to-gas mass
ratio; if no information on the composition is available they are assumed to
be a mixture of graphite and silicates.}
\item{The filling factor, describing the small scale clumpiness of the gas,
can be fixed at any value. If no information is available it is taken to be
unity.}
\item{The distance to the nebula is fixed by an independent individual or statistical method.}
\end{enumerate}

The above assumptions leave the following free parameters: the stellar
temperature, the luminosity of the central star, the hydrogen density in the
ionized region, the inner radius of the nebula, the dust to gas ratio, and the
abundances in the nebula. \\ The outer radius of the nebula is not fixed as an
input parameter, but calculated from the long wavelength end of the dust
emission, or, as a fail-safe, when the electron density drops below
0.1~cm$^{-3}$.

Adopting certain values for the input parameters, it is possible to calculate
a model for the nebula with \cld, predicting the continuum and line fluxes,
photometric magnitudes (including the contribution of line emission) and the
\str\ radius.

To compare the model predictions with the observed quantities, a
goodness-of-fit estimator is calculated. This estimator is minimized by
varying all the input parameters of the model, using the algorithm \amoeba\
(Press et~al. 1986).

It is assumed that there exists a unique set of input parameters, for which
the resulting model predictions give the best fit to a given set of
observables. These input parameters are then considered the best estimate for
the physical properties of the PN.

The full set of observed quantities necessary to derive the physical
parameters of a PN are:
\begin{enumerate}
\renewcommand{\theenumi}{(\arabic{enumi})}
\item The emission line spectrum of the nebula.
Usually this is an optical spectrum, but might also be an ultraviolet and/or
infrared spectrum.  The line ratios make it possible to constrain the
stellar temperature, the density and the electron temperature in the
nebula. They are also required to determine the abundances.  For elements for
which no lines are available we assume standard abundances (Aller \& Czyzak
1983).
\item Since dust is included in the model we also need information on the mid-
and far-infrared continuum. For this the \iras\ fluxes are used.
\item To constrain the emission measure, either an optically thin
radio continuum measurement (e.g. at 6~cm) is needed, or the absolute flux value
of some hydrogen recombination line (usually \hb).  
\item An accurate angular diameter $\Theta_{\rm d}$ of the nebula is needed,
which we define as $\Theta_{\rm d} = 2r_{\rm str}/D$. Here $r_{\rm str}$ stands
for the \str\ radius of the nebula and $D$ is the distance to the nebula.
\end{enumerate}

\section{The sample of galactic bulge PNe}
\label{sample}

We selected a small sample of galactic bulge nebulae from Ratag et al.\ (1997,
RPDM). Galactic bulge nebulae can be assumed to be all at a distance of
approximately 7.8~kpc (Feast 1987). We chose the nebulae from RPDM since they
publish good quality spectra and also carried out their own photo-ionization
analysis of the data which we can use for comparison. The radio observations
for these PNe are described in Gathier et al.\ \cite{c2:gathier}. The
following selection criteria were used:
\begin{enumerate}
\renewcommand{\theenumi}{(\arabic{enumi})}
\item
The PNe should have a quality 2 or 3 \iras\ 12~\mic\ flux and quality 3 \iras\
25~\mic\ and 60~\mic\ fluxes.
\item
The absolute value for the radial velocity should be larger than 100~\kms.
\item
The excitation class should not be labelled peculiar.
\end{enumerate}
The resulting five PNe are presented in \tbl~\ref{ratag:sam}.  All nebulae
except M~2$-$4 are indicated by Acker et al.\ \cite{c2:ack:cat} as likely
bulge PNe.  In view of the large radial velocity of M~2$-$4, $v_{\rm LSR}$ =
$-$175.8~\kms\ (Gathier et al. 1983) it is unlikely to be a foreground
object.

\begin{table}
\caption{Our sample of PNe selected from Ratag et al.\ \protect\cite{c2:ratag}.}
\label{ratag:sam}
\begin{tabular}{ll@{\hspace{6mm}}r@{\hspace{2mm}}r@{\hspace{2mm}}r@{}lr@{\hspace{2mm}}r@{\hspace{2mm}}r}
\hline
 & & \ctr{4}{\hspace{-3mm}$\alpha$(2000)}{0} & \ctr{3}{$\delta$(2000)}{-3} \\
\ctr{1}{name}{0} & \ctr{1}{PN G}{6} & h & m & s && \degr & \arcmin & \arcsec \\
\\
H~1$-$40 & 359.7$-$02.6 & 17 & 55 & 36&.0 & $-30$ & 33 & 33 \\
M~1$-$20 & 006.1$+$08.3 & 17 & 28 & 57&.5 & $-19$ & 15 & 53 \\
M~2$-$4  & 349.8$+$04.4 & 17 & 01 & 06&.2 & $-34$ & 49 & 39 \\
M~2$-$23 & 002.2$-$02.7 & 18 & 01 & 42&.6 & $-28$ & 25 & 44 \\
M~3$-$15 & 006.8$+$04.1 & 17 & 45 & 31&.6 & $-20$ & 58 & 02 \\
\hline
\end{tabular}
\end{table}

\section{Modelling results}
\label{results}

In \tbl~\ref{ratag:inp} we give the input values for the observables used for
the modelling, together with the resulting model predictions. As can be seen
from this table, not all the lines present in the spectra are predicted by
\cld, most notably the higher Balmer lines of hydrogen and several helium
lines. Also the element chlorine is not included in the code. The resulting
physical parameters for the nebulae are given in \tbl~\ref{ratag:phys}. The
hydrogen density shown in this table is the constant density within the \str\
sphere.

\begin{table*}
\small
\caption{Comparison of the observed quantities (mostly taken from Ratag et al.
1997) and the model fit for our sample of PNe. The strength of the emission
lines is given relative to \hb\ = 100. The measured line fluxes have been
dereddened. The entries \fb{O}{ii} \w 3727, \w 7325 and \fb{S}{ii} \w 4071 all
stand for the entire multiplet. All observables for which entries in both
columns obs.\ and model are present, have been weighted in the goodness-of-fit
estimator, except where indicated.}
\label{ratag:inp}
\begin{tabular}{l@{}r @{\hspace{6mm}} r@{.}l@{\x{1.5}}r@{.}l@{}r@{.}l@{\x{1.5}}r@{.}l@{}r@{.}l@{\x{1.5}}r@{.}l@{}r@{.}l@{\x{1.5}}r@{.}l@{}r@{.}l@{\x{1.5}}r@{.}l}
\hline
\ctr{1}{ion}{4} & \ctr{1}{\w}{6} & \ctr{4}{H~1$-$40}{8} & \ctr{4}{M~1$-$20}{7} & \ctr{4}{M~2$-$4}{6} & \ctr{4}{M~2$-$23}{6} & \ctr{4}{M~3$-$15}{2} \\
 & \ctr{1}{\AA}{6} & \ctr{2}{\x{-0.5}obs.}{0} & \ctr{2}{\x{-1}model}{8} & \ctr{2}{\x{-1.5}obs.}{0} & \ctr{2}{\x{-1}model}{7} & \ctr{2}{\x{-1}obs.}{0} &
 \ctr{2}{\x{-1.5}model}{6} & \ctr{2}{\x{0.5}obs.}{0} & \ctr{2}{\x{-1}model}{7}
& \ctr{2}{\x{-1}obs.}{0} & \ctr{2}{\x{-3}model}{0} \\
\hline
\fb{O}{ii}              & 3727 &   32&4\U&  38&5 &   55&5  & 83&1  &     97&1  & 125&2 &      14&2  &  20&4 &   48&6  & 101&5     \\
H\,12                   & 3750 &    \bl  &   \bl &   \bl   & \bl   &     \bl   & \bl   &       3&4  &   \bl &   \bl   & \bl       \\
H\,11,\al{O}{iii}       & 3771 &    \bl  &   \bl &   \bl   & \bl   &     \bl   & \bl   &       4&2  &   \bl &   \bl   & \bl       \\
H\,10                   & 3798 &    \bl  &   \bl &   5&5   & \bl   &     4&0   & \bl   &       5&1  &   \bl &   \bl   & \bl       \\
H\,9                    & 3835 &    \bl  &   \bl &   6&7   & \bl   &     7&1   & \bl   &       6&5  &   \bl &   \bl   & \bl       \\
\fb{Ne}{iii}            & 3869 &   79&1  &  79&4 &   69&1  & 67&6  &     64&7  & 67&6  &      82&4  &  80&0 &   89&9  & 90&1      \\
H\,8,\al{He}{i}         & 3889 &   12&9  &   \bl &   18&1  & \bl   &     17&4  & \bl   &      13&9  &   \bl &   16&6  & \bl       \\
\fb{Ne}{iii},\he        & 3969 &   19&9  &   \bl &   25&5  & \bl   &     33&6  & \bl   &      38&3  &   \bl &   25&4  & \bl       \\
\al{He}{i}              & 4026 &    \bl  &   \bl &   2&7   & \bl   &     2&7   & \bl   &       1&82 &   \bl &   \bl   & \bl       \\
\fb{S}{ii}              & 4071 &    3&3  &   2&5 &   \bl   & 2&3   &     3&6   & 3&0   &      2&5   &  3&8  &   \bl   & 1&8       \\
\hd,\al{N}{iii}         & 4102 &   25&8  &  29&4 &   25&6  & 30&3  &     24&7  & 30&5  &      24&0  &  30&2 &   26&1  & 29&5      \\
\al{C}{ii}              & 4267 \\
\hg                     & 4340 &   47&1  &  50&2 &   49&1  & 51&1  &     45&9  & 51&2  &      48&2  &  50&9 &   49&0  & 50&2      \\
\fb{O}{iii}             & 4363 & 4&6\mm  &  12&0 &   7&5   & 5&4   &     2&9   & 2&1   &      13&9  &   9&9 &   3&3\U & 7&5       \\
\al{He}{i}              & 4472 &    6&5  &   4&4 &   5&8   & 5&2   &     \bl   & 4&5   &       5&1  &   5&5 &   5&0   & 5&1       \\
\al{N}{iii}             & 4641 &    \bl  &   \bl &   \bl   & \bl   &     \bl   & \bl   &       \bl  &   \bl &   20&3  & \bl       \\
\al{He}{ii}             & 4686 & 17&1?\mm&   1&9 &   \bl   & 0&4   &     \bl   & 0&14  &       \bl  &   0&6 &   4&1   & 3&5       \\
\fb{Ar}{iv},\al{He}{i}  & 4712 &    \bl  &   \bl &   1&33  & \bl   &     0&91  & \bl   &      1&09  &   \bl &   1&94  & \bl       \\
\fb{Ar}{iv}             & 4740 &    4&45 &  4&70 &   0&71  & 1&16  &     0&33\U&  0&69 &      0&80  &  1&42 &   \bl   & 6&67      \\
\hb                     & 4861 &   100&  &  100& &   100&  & 100&  &     100&  & 100&  &      100&  &  100& &   100&  & 100&      \\
\al{He}{i}              & 4922 &   1&36\U&   \bl &   \bl   & \bl   &     1&42  & \bl   &      0&87  &   \bl &   \bl   & \bl       \\
\fb{O}{iii}             & 4959 &   307&  &  276& &   336&\m&  315& &    272&\m & 213&  &      304&  &  350& &   328&\m&  219&     \\
\fb{O}{iii}             & 5007 &   915&  &  827& &  1009&\m&  946& &    815&\m & 640&  &     1006&  & 1051& &   983&\m&  658&     \\
\fb{N}{i}               & 5201 &    \bl  &  0&20 &   \bl   & 0&07  &     \bl   & 0&22  &       \bl  &  0&00 &   \bl   & 0&59      \\
\fb{Cl}{iii}            & 5517 &    \bl  &   \bl &   \bl   & \bl   &     \bl   & \bl   &      0&21  &   \bl &   \bl   & \bl       \\
\fb{Cl}{iii}            & 5538 &    \bl  &   \bl &   0&47  & \bl   &     \bl   & \bl   &      0&25  &   \bl &   \bl   & \bl       \\
\fb{N}{ii}              & 5755 &   1&92\U&  2&16 &   0&92  & 1&05  &     1&39  & 1&52  &      1&19  &  1&06 &   1&0\U & 1&4       \\
\al{He}{i}              & 5876 &   15&5  &  15&9 &   15&7  & 16&0  &     13&8  & 13&8  &      17&2  &  17&8 &   16&3  & 16&3      \\
\fb{O}{i}               & 6300 &    2&8  &   2&7 &   4&7   & 4&1   &     2&9   & 3&7   &       4&2  &   3&3 &   2&8   & 3&6       \\
\fb{S}{iii}             & 6312 &   1&70  &  1&48 &   0&79\U&  1&12 &     1&59  & 1&33  &       2&4  &   2&3 &   1&24  & 1&00      \\
\fb{O}{i}               & 6364 &   0&88  &  0&88 &   1&46  & 1&35  &     1&00  & 1&23  &      1&42  &  1&09 &   0&79  & 1&19      \\
\fb{N}{ii}              & 6548 &    \bl  &  19&9 &   \bl   & 14&6  &     \bl   & 32&1  &       \bl  &   6&4 &   \bl   & 18&8      \\
\ha                     & 6563 &   280&  &  278& &   303&  & 269&  &     275&  & 268&  &      283&  &  269& &   305&  & 278&      \\
\fb{N}{ii}              & 6584 &   61&4  &  59&7 &   45&4  & 43&8  &     85&6  & 96&2  &      18&9  &  19&2 &   57&6  & 56&3      \\
\al{He}{i}              & 6678 &    3&8  &   \bl &   4&0   & \bl   &     3&2   & \bl   &       4&7  &   \bl &   4&6   & \bl       \\
\fb{S}{ii}              & 6716 &   0&99  &  0&96 &   1&17  & 1&45  &     2&72  & 3&16  &      0&78  &  0&47 &   2&65  & 3&38      \\
\fb{S}{ii}              & 6731 &   1&71  &  1&99 &   2&32  & 2&91  &     5&0   & 6&0   &      1&55  &  1&05 &   5&4   & 5&2       \\
\al{He}{i}              & 7065 &    7&9  &  11&4 &   10&5  &  9&7  &     5&6   & 7&1   &      14&4  &  12&0 &   7&5   & 9&2       \\
\fb{Ar}{iii}            & 7136 &   13&6  &  13&3 &   9&1   & 7&5   &     15&2  & 13&0  &      14&0  &  11&2 &   19&2  & 19&3      \\
\al{He}{i}              & 7281 &    \bl  &   \bl &   0&83  & \bl   &     0&3\U & \bl   &      0&94  &   \bl &   0&59  & \bl       \\
\fb{O}{ii}              & 7325 &    9&2  &  14&2 &   14&9  & 17&1  &     8&3   & 14&0  &      19&3  &  21&4 &   6&6   & 8&8       \\
\\
\ctr{1}{obs.}{2} & unit \\
\\
$F_\nu$(12 \mic)        & Jy   &    2&38 &  2&38 &    1&13 &  1&00 &      0&56 &  0&53 &      1&93  &  2&10 & $<0$&53\mmm &  0&19     \\
$F_\nu$(25 \mic)        & Jy   &   18&45 & 19&11 &    3&94 &  4&44 &      5&00 &  5&83 &      9&31  &  6&54 &    5&66 &  6&02     \\
$F_\nu$(60 \mic)        & Jy   &   11&91 & 11&42 &    2&38 &  2&30 &      5&77 &  5&18 &      1&64  &  1&64 &    8&02 &  7&77     \\
$F_\nu$(100 \mic)       & Jy   &$<73$&48 &  3&22 & $<4$&59 &  0&66 &  $<12$&59 &  1&73 &  $<126$&70 &  0&24 & $<10$&39&  2&72     \\
$F_\nu$(6 cm)           & mJy  &     31& &  31&0 &     47& &  47&7 &       32& &  32&2 &       41&  &  41&5 &     65& &  65&4     \\
$\Theta_{\rm d}$    &  arcsec  &    1&26 &  1&27 &    1&98 &  1&81 &      2&16 &  2&13 &      0&72  &  0&67 &    5&4  &  5&19     \\
\\
$\chi^2$                &      &     \bl &  0&63 &    \bl  &  1&68 &      \bl  &  2&28 &       \bl  &  4&17 &    \bl  &  2&75     \\
\hline
\end{tabular}
\ntd{\U\x{4}A colon indicates that the value is uncertain.}
\ntd{\m\x{4}The sum of the intensities of the doublet was split using the
ratio 3:1.}
\ntd{\mm\x{4}This line was not weighted in the goodness-of-fit estimator
$\chi^2$, see also \sct~\ref{remarks}.}
\ntd{\mmm\x{4}This flux is not listed as an upper limit in the \iras\
Point Source Catalog, see also \sct~\ref{remarks}.}
\normalsize
\end{table*}

\section{Individual remarks}
\label{remarks}

The PNe in our sample all have nearly the same medium excitation class. This
probably is partially a result of our selection criterion that the nebulae
should have been detected by \iras\ in the 12~\mic\ band (criterion 1). Old
bulge PNe, having a high excitation class and cool dust, might have
insufficient 12~\mic\ flux to be detected by \iras.

In the rest of this section each of the PNe in our sample will be discussed
individually, with special emphasis on the problems encountered during the
modelling.

\subsection{H~1$-$40}
\label{pni:ind}

Two lines were omitted from the list of observables because of the following
reasons. First the \al{He}{ii} \w 4686 line was omitted, because the flux
ratio given by RPDM is quite high, indicative of a high stellar temperature.
However, the rest of the observational data are not consistent with such a
high stellar temperature. Also, this line is listed in \tbl~3 of RPDM, but is
not present in their \tbl~1. Webster (1988, W88) took a spectrum of this PN,
and she didn't report the detection of this line. She should however have
detected a line of the strength mentioned by RPDM. Tylenda et al. \cite{ty94}
list an upper limit of 5 for the intensity of this line. In view of these
uncertainties we decided to omit this line. Since RPDM included this line in
their modelling, this probably explains the higher stellar temperature they
obtain.

The fitting of the \fb{O}{iii} \w 4363 line was also problematic. The observed
flux was far too low to be consistent with the electron temperature predicted
by our model. Since the electron temperature derived from the \fb{N}{ii} line
ratio is much higher (and more consistent with the value determined by our
model), and also because the \fb{O}{iii} \w 4363 line is much stronger in the
spectrum of W88 (however not as strong as predicted by our model), we decided
that its value was too uncertain and omitted it from the input.

\begin{table}
\footnotesize
\caption{The physical parameters of the galactic bulge PNe in our sample
determined with \cld. Abundances of elements for which only one line was
observed are marked uncertain. Since we only model the core region of
M~2$-$23, no values for the outer radius and total shell mass are entered}
\label{ratag:phys}
\begin{tabular}{l@{\x{2.0}}r@{\x{2.0}}r@{\x{2.0}}r@{\x{2.0}}r@{\x{2.0}}r}
\hline
                         & H~1$-$40 & M~1$-$20 & M~2$-$4 & M~2$-$23 & M~3$-$15 \\
\\
$\lg(T_\ast$/K)            &   4.800 &   4.774 &   4.705 &   4.782 &   4.916 \\
$\lg(L_\ast$/\zl)          &   3.798 &   3.607 &   3.555 &   3.639 &   3.663 \\
$\lg(n_{\rm H}$/cm$^{-3}$) &   4.321 &   4.124 &   3.923 &   4.855 &   3.527 \\
$r_{\rm in}$/mpc           &      13 &    0.21 &      11 &       9 &      33 \\
$r_{\rm str}$/mpc          &      24 &      34 &      40 &      13 &      98 \\
$r_{\rm out}$/mpc          &   280\U &   360\U &   350\U &         &   520\U \\
$M_{\rm ion}$/\zm          &   0.042 &   0.092 &   0.088 &   0.015 &    0.47 \\
$M_{\rm sh}$/\zm           &   1.3\U &   2.3\U &   1.9\U &         &   6.5\U \\
$\lg\Gamma$                & $-1.70$ & $-3.11$ & $-2.50$ & $-2.46$ & $-2.60$ \\
$\epsilon$(He)             &   10.96 &   11.02 &   10.96 &   11.05 &   11.03 \\
$\epsilon$(N)              &    7.78 &    7.81 &    8.13 &    7.67 &    7.60 \\
$\epsilon$(O)              &    8.23 &    8.72 &    8.84 &    8.67 &    8.22 \\
$\epsilon$(Ne)             &  7.39\U &  7.82\U &  8.15\U &  7.75\U &  7.53\U \\
$\epsilon$(S)              &    6.37 &    6.66 &    6.90 &    6.79 &    6.31 \\
$\epsilon$(Ar)             &    5.98 &    5.99 &    6.36 &    6.08 &  6.20\U \\
$T_{\rm e}$/kK             &    12.7 &     9.5 &     8.3 &    10.2 &    12.0 \\
$\lg U$                    & $-1.40$ & $+2.14$ & $-1.16$ & $-1.80$ & $-1.58$ \\
\hline
\end{tabular}
\nt{\U\x{4}A colon indicates that the value is uncertain.}
\normalsize
\end{table}

\subsection{M~1$-$20}
\label{pnii:ind}

The intensity of the \ha\ line seems quite high, and is not fitted well. The
discrepancy is too large to be attributed to measurement errors, hence this
might indicate that the spectrum has not been sufficiently dereddened. There
is however no evidence from the fits to the other lines to support this
suspicion.

Our model gives a very small inner radius, also resulting in a very high
ionization parameter. This is caused by the high \iras\ 12~\mic\ over 25~\mic\
flux ratio, which might indicate the presence of hot dust. See also the
discussion in Paper I.

\subsection{M~2$-$4}
\label{pniii:ind}

The spectrum is fitted quite well, but there is slight discrepancy for the
\fb{O}{iii} \w 4959 and \w 5007 lines. This is caused by the \fb{O}{ii} \w
3727 doublet, which is not fitted well. The latter doublet usually has a
larger uncertainty due to extinction and detector insensitivity.

\subsection{M~2$-$23}
\label{pniv:ind}

This PN has the highest $\chi^2$ of all PNe in our sample. This is mainly
caused by the weak lines, which might indicate that this spectrum has a lower
signal-to-noise when compared to the other spectra. RPDM do not list error
margins for their line flux ratios, so we had to assume reasonable values.

The model is not able to fit the \iras\ 25~\mic\ flux, which is very high
compared both to the 12~\mic\ and 60~\mic\ flux. A possible explanation could
be the presence of a 30~\mic\ dust feature in the spectrum (Hoare 1990). This
would imply that the nebula is carbon-rich, since this feature has only been
observed in carbon-rich nebulae. The central star has spectral type Of (Aller
\& Keyes 1987, AK87).

The large difference between the optical diameter of 8.5\arcsec\ (Acker et al.
1992) and the radio diameter of 0.72\arcsec\ (Gathier et al. 1983) suggests
that this nebula might be a core-halo nebula. All other evidence gathered in
this paper also is consistent with this assumption and we will adopt it
throughout the paper. Since we used the radio diameter for the modelling, our
model is only valid for the core region. The fact that our model is density
bounded and gives a low ionized mass is consistent with the fact that we are
only modelling the core region.

\subsection{M~3$-$15}
\label{pnv:ind}

There is a suggestion of a systematic trend when comparing the observed and
the modelled line flux as a function of wavelength. Also the observed
intensity of the \ha\ line seems quite high. This might indicate that the
spectrum has not been sufficiently dereddened.

This PN has a [WC]-type central star (AK87). The central star temperature, the
excitation class 5.5 (taken from RPDM) and the low \iras\ 12~\mic\ to 25~\mic\
flux ratio all are consistent with an early spectral type: [WC3-4] (cf.\ Kaler
1989, M\'{e}ndez \& Niemela 1982, and Zijlstra et al. 1994, respectively).

The \iras\ 12~\mic\ flux is not listed as an upper limit in the Point Source
Catalogue. However, when we used this value for the modelling, the resulting
model was unrealistic. We therefore assume that the 12~\mic\ flux suffers from
confusion and took the quoted value to be an upper limit. See also the
discussion in Paper I.

\section{Discussion}
\label{discussion}

In this section the modelling results are discussed by comparing them with
results from other studies in the literature. Since distance dependent
parameters are usually not given by other authors, we will restrict ourselves
to the distance independent parameters of PNe.

\subsection{Stellar temperatures}

In \tbl~\ref{temp:tab} we present a comparison with the stellar temperatures
given in the literature.  These temperatures were derived using the Zanstra
method and photo-ionization modelling. Results using the energy balance or
Stoy method are not listed since we consider this method, or at least the
data for the nebulae being studied here, to be unreliable (Pottasch, private
communication). One can see that the derived values agree quite well with only
a few outliers.

The temperatures determined by our method agree well with the hydrogen Zanstra
temperatures, with the single exception of the temperature for M~2$-$23 given
by Tylenda et al.\ \cite{c2:tylenda}. Since the other three determinations
using the Zanstra method agree quite well, we assume the value given by
Tylenda et al.\ \cite{c2:tylenda} to be erroneous.

\begin{table}
\caption{Comparison of the stellar temperatures for the PNe in our sample. The
temperatures are given in kilokelvin. The abbreviations for the methods have
the following meaning: \al{H}{i}~-- hydrogen Zanstra method, \al{He}{ii}~--
helium Zanstra method, AM~-- photo-ionization modelling using model
atmospheres, BB~-- photo-ionization modelling using blackbody approximation.}
\label{temp:tab}
\begin{tabular}{r@{\x{5}}r@{\x{5}}r@{\x{5}}r@{\x{5}}r@{\x{5}}r@{\x{2}}r}
\hline
H~1$-$40\hspace*{-2.5mm} & M~1$-$20\hspace*{-2.5mm} & M~2$-$4\hspace*{-2.5mm} & M~2$-$23\hspace*{-2.5mm} & M~3$-$15\hspace*{-2.5mm} & ref. & meth.\\
\\
             &      & 55.\f& 64.\f&      & 3 & \al{H}{i} \\
             & 53.\f&      & 56.\f&      & 4 & \al{H}{i} \\
             &      & 51.5 & 65.0 &      & 6 & \al{H}{i} \\
             & 65.\f&      & 85.\f&      & 8 & \al{H}{i} \\
             & 49.5 & 49.9 &      &      & 5 & \al{H}{i} \\
             & 59.9 &      &      &      & 5 &\al{He}{ii}\\
             &      &      & 50.\f& 62.5 & 1 &        AM \\
 80.0\rlap{?}& 50.0 & 50.0 & 57.5 & 72.5 & 7 &        AM \\
 64.\f       &      &      &      &      & 2 &        BB \\
 63.1        & 59.4 & 50.7 & 60.5 & 82.4 & 9 &        BB \\
\hline
\end{tabular}
\nt{References ---
1. Aller \& Keyes \cite{c2:aller:keyes} using model atmospheres by Husfeld et al.\ \cite{c2:husfeld}
2. Dopita et al.\ \cite{c2:dopita}
3. Gleizes, Acker \& Stenholm \cite{c2:gleizes}
4. Kaler \& Jacoby \cite{c2:kej}
5. Mal'kov \cite{malkov}
6. Pottasch \& Acker \cite{c2:pot:ack}
7. Ratag et al.\ \cite{c2:ratag} using model atmospheres by Clegg \& Middlemass \cite{c2:cem} and Husfeld et al.\ \cite{c2:husfeld}
8. Tylenda et al.\ \cite{c2:tylenda}
9. This work}
\end{table}

To derive stellar temperatures for photo-ionization modelling, sometimes
certain line-ratios are used as temperature indicators (e.g.\ \al{He}{ii} \w
4686 over \hb). Especially for cooler central stars, where few temperature
sensitive lines are available, this makes the determination dependent on one
or two lines. Nevertheless, the results from the photo-ionization models
usually are in good agreement. Exceptions are the temperature for H~1$-$40
derived by RPDM, and the temperatures for M~3$-$15. The deviating value for
H~1$-$40 given by RPDM can probably be attributed to the \al{He}{ii} \w 4686
line, which they used as a temperature indicator. We refer to the discussion
in \sct~\ref{pni:ind}. For M~3$-$15 we find a higher stellar temperature than
other authors. The largest discrepancy is with the value from AK87. This can
probably be attributed to the fact that AK87 did not report a detection of the
\al{He}{ii} \w 4686 line in their spectrum (although a detection of roughly
the same strength as RPDM was reported in Aller \& Keyes 1985). Since M~3$-$15
has a [WC]-type central star, part of the \al{He}{ii} \w 4686 flux may
originate from the central star. Unfortunately, no detection of the
\fb{Ar}{iv} \w 4740 line has ever been reported, so that no alternative
temperature sensitive line is available. In view of this, the central star
temperature for M~3$-$15 should be viewed with some caution.

We conclude that the temperature determination for the central stars in this
sample is fairly reliable, although the situation for M~3$-$15 is not
completely clear. This confirms our results from Paper I in which we found the
temperature determination to be robust.

\subsection{Electron temperatures}
\label{te:litt}

In \tbl~\ref{te:tab} the electron temperatures derived by different authors
are compared. The electron temperature determined by \cld\ is a weighted mean
of the temperature in the nebula: $\overline{T_{\rm e}}$ = $\int n_{\rm e}^2
T_{\rm e} \md V / \int n_{\rm e}^2 \md V$. The observational material shows a
large spread in most cases, even when the same method is used. This indicates
that the electron temperature determination, at least in those cases where
diagnostic lines have been used, is not very reliable. This is in agreement
with our results in Paper I. Note the large difference between the \fb{N}{ii}
and \fb{O}{iii} temperatures in the case of M~2$-$23. This difference is not
caused by measurement error. For this particular object, the temperature
derived from the \fb{N}{ii} lines has no physical meaning (Liu, private
communication).

\begin{table}
\caption{Various determinations of the electron temperature for the nebulae in
our sample. The temperatures are given in kilokelvin. The abbreviations for
the methods have the following meaning: ave.~-- average of \fb{N}{ii} and
\fb{O}{iii}, model~-- average model prediction (see text).}
\label{te:tab}
\begin{tabular}{r@{\x{5}}r@{\x{5}}r@{\x{5}}r@{\x{5}}r@{\x{5}}r@{\x{2}}r}
\hline
H~1$-$40\hspace*{-2.5mm} & M~1$-$20\hspace*{-2.5mm} & M~2$-$4\hspace*{-2.5mm} & M~2$-$23\hspace*{-2.5mm} & M~3$-$15\hspace*{-2.5mm} & ref. & meth.\\
\\
     &      &      & 18.0 &      & 1  &  \fb{N}{ii} \\
     & 17.1 &      &      &      & 2  &  \fb{N}{ii} \\
     & 12.5 & 12.5 &      &      & 4  &  \fb{N}{ii} \\
     &      & 10.2 &      &      & 7  &  \fb{N}{ii} \\
     & 10.4 &      &      &      & 8  &  \fb{N}{ii} \\
13.1 & 10.2 &  9.7 & 19.2 & 9.4\U& 9  &  \fb{N}{ii} \\
10.1 &      &      &      &      & 11 &  \fb{N}{ii} \\
     &      &      & 11.0 &      & 1  & \fb{O}{iii} \\
     & 10.8 &      &      &      & 2  & \fb{O}{iii} \\
     & 13.1 & 11.6 &      &      & 4  & \fb{O}{iii} \\
     &      &  8.7 &      &      & 5  & \fb{O}{iii} \\
     &      &      & 12.9 &      & 6  & \fb{O}{iii} \\
     &  9.9 &      &      &      & 8  & \fb{O}{iii} \\
 9.3 & 10.4 &  8.5 & 13.0 & 8.4\U& 9  & \fb{O}{iii} \\
 9.7 &      &      &      &      & 11 & \fb{O}{iii} \\
     &      &      & 12.6 &11.2\U& 3  &        ave. \\
     &      & 11.1 &      &      & 10 &        ave. \\
12.7 &  9.5 &  8.3 & 10.2 & 12.0 & 12 &       model \\
\hline
\end{tabular}
\nt{\U\x{4}A colon indicates that the value is uncertain.}
\nt{References ---
1. Acker et~al.\ \cite{c2:ackko}
2. Acker et~al. \cite{c2:ackko2}
3. Aller \& Keyes \cite{c2:aller:keyes}
4. Costa et al.\ \cite{c2:costa}
5. Cuisinier, Acker \& K\"oppen \cite{c2:cuisinier}
6. Kaler \cite{c2:kaler:i}
7. Kaler et al. \cite{c2:kalerea}
8. Kaler et al. \cite{c2:kalerea2}
9. Ratag et al.\ \cite{c2:ratag}
10. Tylenda et al. \cite{c2:tylenda2}
11. Webster \cite{c2:webster}
12. This work}
\end{table}

The electron temperatures derived from our method are in most cases just
outside the range of values found with line diagnostics; three times at the
low end and twice at the high end. The results from Paper~I indicate that the
electron temperature determination with our method should be robust. It is not
apparent to us why the average values of the electron temperature derived from
line diagnostics do not coincide with our results. This might indicate a
problem, although the fact that we find both higher and lower results is not
indicative of a systematic effect. Nevertheless, this issue should be
investigated further in future research, using a larger sample.

\subsection{Electron densities}
\label{eden:disc}

In \tbl~\ref{ne:tab} the electron densities derived by different authors are
compared. The electron density determined by \cld\ is a weighted mean of the
density in the nebula: $\overline{n_{\rm e}}$ = $\int n_{\rm e}^3 \md V / \int
n_{\rm e}^2 \md V$. There are enormous differences between the various
determinations in the literature, even when the same method has been applied.
This indicates that the determination of densities with line diagnostics is
unreliable, which confirms our results in Paper~I. Also note the enormous
differences between the \fb{S}{ii}, \fb{Cl}{iii} and \fb{Ar}{iv} densities for
M~1$-$20 derived by Kaler et al. \cite{c2:kalerea2}.

\begin{table}
\caption{Various determinations of the electron density for the nebulae in our
sample. The densities are given in $10^3$~cm$^{-3}$. The abbreviations for the
methods have the following meaning: radio~-- density determined from the radio
flux, model~-- average model prediction (see text).}
\label{ne:tab}
\begin{tabular}{r@{\x{5}}r@{\x{5}}r@{\x{5}}r@{\x{5}}r@{\x{5}}r@{\x{2}}r}
\hline
H~1$-$40\hspace*{-2.5mm} & M~1$-$20\hspace*{-2.5mm} & M~2$-$4\hspace*{-2.5mm} & M~2$-$23\hspace*{-2.5mm} & M~3$-$15\hspace*{-2.5mm} & ref. & meth.\\
\\
      &      &      & 13.6 &      & 1  &   \fb{S}{ii} \\
 10.9 &  4.7 &      &      & 24.3 & 2  &   \fb{S}{ii} \\
      &      &      &  3.0 &  2.5 & 3  &   \fb{S}{ii} \\
      & 17.8 &  4.5 &      &      & 5  &   \fb{S}{ii} \\
      &      &  7.0 &      &      & 6  &   \fb{S}{ii} \\
 15.0 &      &      &      &      & 7  &   \fb{S}{ii} \\
      &      &  4.2 &      &      & 9  &   \fb{S}{ii} \\
      & 85.\f&      &      &      & 10 &   \fb{S}{ii} \\
  4.4 &  9.2 &  5.6 & 11.5 & 10.6 & 11 &   \fb{S}{ii} \\
      &      &      &      &  4.2 & 13 &   \fb{S}{ii} \\
      &      &  3.6 &      &      & 14 &   \fb{S}{ii} \\
 35.1 &      &      &      &      & 16 &   \fb{S}{ii} \\
      &      &  5.7 &      &      & 9  & \fb{Cl}{iii} \\
      &  7.8 &      &      &      & 10 & \fb{Cl}{iii} \\
      &      &      & 79.\f&      & 4  &  \fb{Ar}{iv} \\
      &  1.0 &      &      &      & 10 &  \fb{Ar}{iv} \\
      &      &      & 63.\f&      & 15 &  \fb{Ar}{iv} \\
 13.5 &      &      &      &      & 7  &        radio \\
      &      &      & 20.\f&      & 8  &        radio \\
      & 10.\f&      &      &      & 12 &        radio \\
 22.7 & 14.6 &  9.1 & 79.3 &  3.7 & 17 &        model \\
\hline
\end{tabular}
\nt{References ---
1. Acker et al.\ \cite{c2:ackko}
2. Acker et al. \cite{c2:ackko2}
3. Aller \& Keyes \cite{c2:aller:keyes}
4. Boffi \& Stanghellini \cite{c2:boffi}
5. Costa~et al.\ \cite{c2:costa}
6. Cuisinier et al. \cite{c2:cuisinier}
7. Dopita et al.\ \cite{c2:dopita}
8. Kaler \cite{c2:kaler:i}
9. Kaler et al. \cite{c2:kalerea}
10. Kaler et al. \cite{c2:kalerea2}
11. Ratag et al.\ \cite{c2:ratag}
12. Shaw \& Kaler \cite{c2:shaw2}
13. Stanghellini \& Kaler \cite{c2:stang}
14. Tylenda et al. \cite{c2:tylenda2}
15. Webster \cite{c2:web76}
16. Webster \cite{c2:webster}
17. This work}
\end{table}

Our values differ substantially from the values given by RPDM, although they
are based on the same observational data. This is because we use a completely
different method to determine the density. For three out of five nebulae we
find results which are within the range of values found with other methods.
For M~2$-$4 we find a value which is a bit larger. The results in Paper~I
indicate that our determination of the density is somewhat susceptible to
measurement errors and errors in the model assumptions. This might provide an
explanation for the discrepancy. The fact that we model only the core region
of M~2$-$23 provides an explanation for the very high density we find for this
nebula. Webster \cite{c2:web76} and Boffi \& Stanghellini (1994, using the
same spectrum) also find a high value using the \fb{Ar}{iv} line ratio. The
\fb{Ar}{iv} lines are expected to be formed predominantly in the core region
and hence this would confirm our results. On the other hand, the excitation in
the core is too high to form large amounts of S$^+$. Hence the \fb{S}{ii}
lines can be expected to originate predominantly from the halo and should
therefore indicate lower densities. This of course also depends on the exact
position of the slit over the nebula. All of this might be an explanation for
the extremely large spread of values found for this nebula.

The quality of the data in \tbl~\ref{ne:tab} makes a comparison with our
results meaningless. However, the data are at least consistent with the
assumption that our results are more accurate than the results from line
diagnostics.

\subsection{Nebular abundances}

\begin{table*}
\small
\caption{Comparison of the abundance determinations of the PNe in our sample.}
\label{abun:lit}
\begin{tabular}{lrrrrrrrrrrrrrr}
\hline
 & \ctr{4}{H~1$-$40}{0} & & \ctr{4}{M~1$-$20}{0} & & \ctr{4}{M~2$-$4}{0} \\
ref:\x{5} &\sctr{4}&\sctr{9}&\sctr{12}&\sctr{13} & &\sctr{2}&\sctr{7}&\sctr{9}&\sctr{13} & &\sctr{2}&\sctr{3}&\sctr{9}&\sctr{13}\\
\noalign{\vskip2pt}
\\
\noalign{\vskip2pt}
$\epsilon$(He) & 11.03 & 11.06 & 11.04 & 10.96 &       & 11.07 & 10.94 & 11.02 & 11.02 &       & 11.11 &10.96\U& 10.99 & 10.96 \\
$\epsilon$(N)  &  7.72 &  8.08 &       &  7.78 &       &  7.39 &  7.80 &  7.75 &  7.81 &       &  7.65 &  8.17 &  8.09 &  8.13 \\
$\epsilon$(O)  &  8.52 &  8.70 &  8.53 &  8.23 &       &  8.30 &  8.65 &  8.62 &  8.72 &       &  8.30 &  8.77 &  8.80 &  8.84 \\
$\epsilon$(Ne) &  7.89 &  7.69 &       & 7.39\U&       &       &       &  7.79 & 7.82\U&       &       &       &  7.90 & 8.15\U\\
$\epsilon$(S)  &  6.88 &  6.77 &       &  6.37 &       & 6.43\U&       &  6.52 &  6.66 &       &  6.64 &  7.03 &  6.96 &  6.90 \\
$\epsilon$(Ar) &  6.6\f&  6.43 &       &  5.98 &       &       &       &  6.05 &  5.99 &       &  6.31 &  6.36 &  6.25 &  6.36 \\
\\
 & \ctr{8}{M~2$-$23}{0} & & \ctr{4}{M~3$-$15}{0} \\
ref:\x{5} &\sctr{1}&\sctr{5}&\sctr{6}&\sctr{8}&\sctr{9}&\sctr{10}&\sctr{11}&\sctr{13} & &\sctr{1}&\sctr{5}&\sctr{9}&\sctr{13}\\
\noalign{\vskip2pt}
\\
\noalign{\vskip2pt}
$\epsilon$(He) & 11.00 & 10.93 &       &10.88\U& 10.98 & 10.92 & 10.96 & 11.05 &       & 11.03 & 11.01 & 11.03 & 11.03 \\
$\epsilon$(N)  &  7.68 &  8.20 &  7.70 &  7.55 &  7.40 &  8.13 &       &  7.67 &       &  8.08 &       &  8.14 &  7.60 \\
$\epsilon$(O)  &  8.40 &  8.42 &  8.18 &  8.34 &  8.22 &  8.47 &  8.11 &  8.67 &       &  8.41 &  8.51 &  8.74 &  8.22 \\
$\epsilon$(Ne) &  7.60 &  7.62 &       & 6.46\U&  7.15 &  7.65 &       & 7.75\U&       &  7.48 &  7.62 &  7.86 & 7.53\U\\
$\epsilon$(S)  &  6.6\f&       &       &  6.29 &  6.30 &       &       &  6.79 &       &6.7\U\f&       &  6.86 &  6.31 \\
$\epsilon$(Ar) & 5.75  &       &       &  5.86 &  5.81 &       &       &  6.08 &       &  6.5\f&       &  6.53 & 6.20\U\\
\hline
\end{tabular}
\ntd{\U\x{4}A colon indicates that the value is uncertain.}
\ntd{References ---
1. Aller \& Keyes \cite{c2:aller:keyes}
2. Costa et al. \cite{c2:costa}
3. Cuisinier et al. \cite{c2:cuisinier}
4. Dopita et al.\ \cite{c2:dopita}
5. Henry \cite{c2:henry}
6. Kaler \cite{c2:kaler:ii}
7. Kaler et al. \cite{c2:kalerea2}
8. K\"oppen (private communication)
9. Ratag et al. \cite{c2:ratag}
10. Walton, Barlow \& Clegg \cite{c2:wal:bar}
11. Webster \cite{c2:web76}
12. Webster \cite{c2:webster}
13. This work}
\normalsize
\end{table*}

In \tbl~\ref{abun:lit} we give a comparison of the abundances we determined
with other literature values. We did not include the nitrogen abundance for
M~3$-$15 from Henry \cite{c2:henry}. After a discussion with Dr. Henry it was
established that this abundance was flawed by an error in the analysis (as is
also the case for the nitrogen abundances of M~4$-$3 and H~1$-$23 listed in
the same paper; all other results are not affected). We also did not include
the abundances for M~2$-$23 listed in K\"oppen, Acker \& Stenholm
\cite{c2:koppen}. It was established that this analysis was flawed by an error
as well, and Dr. K\"oppen kindly provided us with a re-analysis of his data.
The higher nitrogen abundance given by Walton et al.\ \cite{c2:wal:bar} for
M~2$-$23 might be a result of the inclusion of \iue\ data in their analysis.
They systematically find higher nitrogen abundances for bulge PNe than other
authors.

One can see that large differences can be found between the various abundance
determinations in the literature. If we exclude our own results, we find the
following statistics. For elements heavier than helium, we find a difference
between the lowest and highest abundance determination larger than or equal to
0.3~dex in 12 out of 22 cases, and larger than or equal to 0.5~dex in 4 out of
22 cases. For the helium abundances we find a spread larger than 0.1~dex in 2
out of 5 cases. Especially the abundances for M~2$-$23 show a large spread and
should be considered uncertain. From this we draw the conclusion that, at
least for the sample studied here, abundance determinations can not be
considered very accurate. Uncertainties exceeding 0.2~dex to 0.3~dex are not
uncommon.

When one compares the abundances for the individual PNe with the values from
RPDM, one can see that for the two objects where the electron temperature is
in good agreement (M~1$-$20 and M~2$-$4), the abundances also agree very well.
For the other objects the abundance determinations differ. We attribute this
to the difference in the determination of the electron temperature. When we
compare our abundance determinations with the other values found in the
literature, we see that our results often are slightly outside the range of
values found by other authors. This behaviour is well correlated: either all
outliers are at the low end or at the high end. This behaviour is also well
correlated with our electron temperature determination. When our electron
temperature determination is at the low end, our abundances are at the high
end, and the reverse is also the case (see also \sct~\ref{te:litt}). This
indicates that the main source of uncertainty in the abundance determination
is the electron temperature. Hence the discussion given in \sct~\ref{te:litt}
applies here as well.

\subsection{Stellar broadband fluxes}

Since \cld\ calculates the attenuation of the stellar continuum separately
from the transport of the diffuse nebular continuum, we are able to predict
broadband photometric fluxes for the central star as they would appear through
the nebula. In this way we could calculate a prediction for the Johnson~\jb\
and \jv\ magnitudes. However, observed stellar magnitudes will be reddened due
to interstellar extinction as well, and we have to take this into account in
our predictions. To calculate the total extinction towards the nebula, we
averaged all the measurements we could find in the literature. Since the
continuum fluxes predicted by \cld\ already take the internal extinction into
account, we only have to correct the stellar magnitudes for the external
extinction. Hence we used the internal extinction from our model, and
subtracted it from the total extinction. Then we used this value for the
external extinction to predict the reddened \jb\ and \jv\ magnitudes of the
central star. Where necessary, we applied the interstellar reddening law given
by Pottasch \cite{c2:book:pot}. A comparison of the calculated values with the
literature values taken from Acker et al.\ \cite{c2:ack:cat} is given in
\tbl~\ref{mag:tab}.

\begin{table*}
\small
\caption{For the PNe in our sample we give in column 2 and 3 the Johnson~\jb\
and \jv\ magnitudes resp.\ predicted by our model, in column 4 the internal
extinction in the Johnson~\jv\ band derived from our model, in column 5 the
average total extinction derived from the Balmer decrement and the radio flux,
in columns 6 and 7 the predicted reddened values for the Johnson~\jb\ and \jv\
magnitudes and in column 8 and 9 the measured magnitudes given in Acker et
al.\ \protect\cite{c2:ack:cat}.}
\label{mag:tab}
\begin{tabular}{lcccccccc}
\hline
name & \jb$_{\rm mod}$ & \jv$_{\rm mod}$ & $A_\jv^{\rm int}$ & $A_\jv^{\rm tot}$ & \jb$_{\rm pred}$ & \jv$_{\rm pred}$ & \jb & \jv \\
     & \mgn & \mgn & \mgn & \mgn & \mgn & \mgn & \mgn & \mgn \\
\\
H~1$-$40 & 15.70 & 15.68 & 1.11 & 5.05$\pm$0.26 & 20.86$\pm$0.34 & 19.62$\pm$0.26 &      &      \\
M~1$-$20 & 14.62 & 14.90 & 0.06 & 2.61$\pm$0.10 & 17.96$\pm$0.13 & 17.45$\pm$0.10 & 17.7 & 17.1 \\
M~2$-$4  & 14.41 & 14.64 & 0.14 & 2.78$\pm$0.15 & 17.87$\pm$0.19 & 17.28$\pm$0.15 & 17.6 & 17.0 \\
M~2$-$23 & 14.63 & 14.90 & 0.08 & 1.81$\pm$0.22 & 16.90$\pm$0.28 & 16.63$\pm$0.22 & 16.7 &      \\
M~3$-$15 & 15.52 & 15.80 & 0.10 & 4.69$\pm$0.25 & 21.53$\pm$0.33 & 20.39$\pm$0.25 &      &      \\
\hline
\end{tabular}
\normalsize
\end{table*}

The predicted magnitudes are slightly fainter than observed, but still in
remarkable good agreement, considering the fact that we use a blackbody
approximation to determine these values. Given the fact that a blackbody of a
given temperature has more ionizing photons than a realistic spectrum with the
same effective temperature, one can expect that in the best-fit model the
total luminosity will be underestimated to compensate for this effect.
However, we find that this effect is only very modest and this can be
understood from the fact that we include the dust emission in the modelling.
Grains can be heated very efficiently by Balmer continuum photons, as well as
by Lyman continuum photons. Therefore, the \iras\ fluxes give a good
constraint on the Balmer continuum flux. This counteracts the previously
mentioned underestimation of the total luminosity and explains the remarkable
accuracy of our stellar broadband fluxes.

\subsection{Distances}
\label{dist:discussion}

In our model assumptions we assume the distance to be a fixed number. However,
our method can easily be changed in such a way that the distance would be a
free parameter. When this is done, the best-fit model would also give an
estimate for the distance. We have investigated the possibility to determine
the distance this way (van Hoof \& Van de Steene 1996). We found that, though
possible in principle, the spread in the resulting distance determinations is
large. The distance determination is vulnerable to various observational
errors, but especially to the error in the determination of the angular
diameter. Since the angular diameter is notoriously hard to measure, this
sensitivity makes the results very uncertain. When we determined the distances
to the bulge nebulae in our sample with this method, we found the spread in
the values to be larger than what is obtained from a statistical method (Van
de Steene \& Zijlstra 1995).

Closer investigation reveals that this method of determining distances is in
essence identical to the method described in Phillips \& Pottasch
\cite{c2:phillips}. They already concluded that this method is unreliable. The
use of a wrong value for the distance not only influences the distance
dependent parameters but also some distance independent parameters, as was
already discussed in Paper~I. We therefore do not recommend this method, and
advise the use of separately determined distances.

\section{Conclusions}
\label{conclusions}

We applied our method which enables a fully self-consistent determination of
the physical parameters of a PN, using the spectrum, the \iras\ and radio
fluxes and the angular diameter of the nebula, to a sample of five galactic
bulge PNe. Comparison of the distance independent physical parameters with
published data shows that the stellar temperatures generally are in good
agreement and can be considered reliable. The literature data for the electron
temperature, electron density and also for the abundances show a large spread,
indicating that the use of line diagnostics is not reliable. Comparison of the
various abundance determinations indicates that the uncertainty in the
electron temperature is the main source of uncertainty in the abundance
determination. The large spread in the literature data makes a comparison with
our results meaningless. The stellar magnitudes predicted by the
photo-ionization models are in good agreement with observed values.

\section*{Acknowledgments}
We thank Drs.\ S.R.\ Pottasch, L.B.F.M.\ Waters and J.M.\ van der Hulst for
helpful comments. Drs.\ R.D.\ Oudmaijer, T. de Jong, G.J. Ferland and the
anonymous referee are thanked for critically reading the manuscript. Drs.
R.B.C. Henry and J. K\"oppen are thanked for re-examining their data. The
authors acknowledge support from the Netherlands Foundation for Research in
Astronomy (ASTRON) through grant no.\ 782--372--033 and 782--372--035. PvH
wishes to thank ESO for their hospitality and financial support during his
stay in Santiago where part of this paper was written. PvH is currently
supported by the NSF through grant no.\ AST 96--17083. The photo-ionization
code \cld\ was used, written by Gary Ferland and obtained from the University
of Kentucky, USA. We thank Gary Ferland for his invaluable help in debugging
and modifying the code.

\end{document}